\documentclass{aa}
\usepackage{graphics,epsfig,txfonts}

\usepackage{natbib}
\bibpunct{(}{)}{;}{a}{}{,}

\newcommand{\Hone}{\ion{H}{i}}

\newcommand{\Halp}{H${\alpha}$}
\newcommand{\ltsima}{$\buildrel < \over \sim$}
\newcommand{\lsim}{\lower.5ex\hbox{\ltsima}}
\newcommand{\gtsima}{$\buildrel > \over \sim$}
\newcommand{\gsim}{\lower.5ex\hbox{\gtsima}}
\newcommand{\xmm}{XMM-{\it Newton}}

\def\rahour{\hbox{\ensuremath{^{\rm h}}}}
\def\ramin{\hbox{\ensuremath{^{\rm m}}}}

\newcommand{\bexrb}{RX\,J0123.4-7321}

\usepackage{color}

\begin{document}
 
\title{RX\,J0123.4-7321, a Be/X-ray binary in the wing of the SMC
      }

\author{      R.~Sturm\inst{1} 
     \and     F.~Haberl\inst{1}  
     \and     W.~Pietsch\inst{1}  
     \and     A.~Udalski\inst{2}  
       }

\titlerunning{The Be/X-ray binary RX\,J0123.4-7321}
\authorrunning{Sturm et al.}

\institute{ Max-Planck-Institut f\"ur extraterrestrische Physik, Giessenbachstra{\ss}e, 85748 Garching, Germany
	    \and
            Warsaw University Observatory, Aleje Ujazdowskie 4, 00-478 Warsaw, Poland
	   }

\date{Received 8 November 2012 / Accepted 13 January 2013}

 \abstract{}
          {To confirm faint Be/X-ray binary candidates from the \xmm\ survey of the Small Magellanic Cloud, we searched for X-ray outbursts in archival ROSAT observations. 
           We found that \bexrb\ was much brighter when detected with ROSAT than seen 16 years later by \xmm.} 
          {We analysed the ROSAT observations and the OGLE $I$-band light curve of the optical counterpart to investigate the nature of the system.}
          {High long-term variability in the X-ray flux of a factor of $\sim$150 was found between the ROSAT and \xmm\ detections, indicating strong outburst activity during the ROSAT observations.
           The $I$-band light curve reveals long-term variability and regular outbursts with a period of (119.9$\pm$2.5) days indicating the orbital period of the binary system.}
          {The large X-ray flux variations and the properties of the optical counterpart confirm \bexrb\ as a new Be/X-ray binary in the wing of the Small Magellanic Cloud.}

%abstract without latex:
%
%To confirm faint Be/X-ray binary candidates from the XMM-Newton survey of the Small Magellanic Cloud, we searched for X-ray outbursts in archival ROSAT observations. We found that RXJ0123.4-7321 was much brighter when detected with ROSAT than seen 16 years later by XMM-Newton. We analysed the ROSAT observations and the OGLE I-band light curve of the optical counterpart to investigate the nature of the system. High long-term variability in the X-ray flux of a factor of ~150 was found between the ROSAT and XMM-Newton detections, indicating strong outburst activity during the ROSAT observations. The I-band light curve reveals long-term variability and regular outbursts with a period of (119.9+-2.5) days indicating the orbital period of the binary system. The large X-ray flux variations and the properties of the optical counterpart confirm RXJ0123.4-7321 as a new Be/X-ray binary in the wing of the Small Magellanic Cloud.

 
\maketitle
 
\section{Introduction}
\label{sec:introduction}

Be/X-ray binaries \citep[BeXRBs, e.g. ][]{2011Ap&SS.332....1R} are binary systems composed of  a compact object and a Be star of spectral type O9$-$B5 and luminosity class III$-$V.
Be stars eject matter in the equatorial plane leading to the build up of a decretion disc \citep{2001PASJ...53..119O,2002MmSAI..73.1038Z} 
around the star that produces emission lines and a near-infrared (NIR) excess.
Both components show variability due to disc instabilities and sometimes due to interaction with the compact object. 
The compact object is in most cases a neutron star (NS).
Systems with black holes are not known up to now. However, systems may contain white dwarfs \citep{2012ApJ...761...99L,2012A&A...537A..76S}
that are observable at SMC distance as super-soft X-ray source in the case of thermonuclear surface burning on the white dwarf.
Caused by a kick experienced by the NS during the supernova explosion, the NS can have an eccentric orbit around the Be star.
Accretion is possible, when the NS is close to the decretion disc of the Be star during periastron passage.
This causes so-called type-I outbursts that last for some days.
Longer (several weeks) type-II outbursts are likely connected with decretion disc instabilities \citep[][]{2002MmSAI..73.1038Z}.

The Small Magellanic Cloud (SMC) hosts a large population of BeXRBs.
A literature search \citep[e.g. ][]{2004A&A...414..667H,2005MNRAS.356..502C,2005A&A...442.1135L,2010ApJ...716.1217L} reveals about 60 pulsars in more than 100 BeXRB systems (including candidates), 
which allows a comparison of their statistical properties with those of the Galactic sample.
Most of the BeXRBs in the SMC are found in the bar where increased star formation $\sim$40 Myr ago created them in large number \citep{2010ApJ...716L.140A}.
However, in the eastern wing of the SMC only a few systems are known, probably due to a younger stellar population \citep[$\sim$10 Myr, ][]{2004AJ....127.1531H}.
Also the BeXRBs in the wing might be younger, as e.g. SXP\,1062 found in a supernova remnant suggests \citep{2012MNRAS.420L..13H,2012A&A...537L...1H}.
This might cause different properties of the wing sample.
E.g. a literature search reveals in the wing area three known BeXRB pulsars with $P_{\rm spin}<40$ s \citep{1997ApJ...474L.111C,1998IAUC.7048....1C,2003IAUC.8064....4C}
compared to only one (SXP\,1062) pulsar with $P_{\rm spin}>40$ s. 
In contrast to that, for the whole SMC \citet{2011Natur.479..372K} find a bimodal spin distribution with more systems above $P_{\rm spin}>40$ s,
which is likely caused by a bimodal distribution of orbital periods.
The search for additional systems in the wing of the SMC is essential to enable a statistical comparison of the wing and bar population and exclude any observational bias.

The \xmm\ survey of the SMC \citep{2012A&A...545A.128H} revealed new BeXRBs in X-ray-bright state \citep{2011MNRAS.410.1813T,2011A&A...527A.131S,2011MNRAS.414.3281C,2012MNRAS.424..282C}.
In addition, several candidates for BeXRBs were found in the \xmm\ SMC point-source catalogue \citep{smcsrccat}.
One of these candidates is \bexrb\ \citep{2000A&AS..142...41H}, which was detected with ROSAT \citep{1982AdSpR...2..241T} previously at much higher X-ray luminosity, but not recognised as BeXRB.
In this paper, we present our analysis of the ROSAT observations and the OGLE \citep{2008AcA....58...69U} data of the optical counterpart,
allowing us to confirm \object{\bexrb} as a new BeXRB in the wing of the SMC with a determined orbital period.

\section{Analyses and results of X-ray observations}
\label{sec:x-ray}

%--------------------------
\begin{table*}
\centering
\caption[]{X-ray observations of \bexrb.}
\begin{tabular}{lcccccrcccc}
\hline\hline\noalign{\smallskip}
\multicolumn{1}{c}{Satellite} &
\multicolumn{1}{c}{ObsID} &
\multicolumn{1}{c}{MJD\tablefootmark{a}} &
\multicolumn{1}{c}{Dur\tablefootmark{b}} &
\multicolumn{1}{c}{Orbit\tablefootmark{c}} &
\multicolumn{1}{c}{Exp} &
\multicolumn{1}{c}{Cts} &
\multicolumn{1}{c}{$F_{\rm observed}$} &
\multicolumn{1}{c}{$F_{\rm model}$} &
\multicolumn{1}{c}{$L$\tablefootmark{d}} &
\multicolumn{1}{c}{$L$\tablefootmark{d}} \\
\multicolumn{1}{c}{Instrument}&
\multicolumn{1}{c}{} &
\multicolumn{1}{c}{} &
\multicolumn{1}{c}{(d)} &
\multicolumn{1}{c}{Phase} &
\multicolumn{1}{c}{(ks)} &
\multicolumn{1}{c}{} &
\multicolumn{1}{c}{( erg cm$^{-2}$ s$^{-1}$)}&
\multicolumn{1}{c}{( erg cm$^{-2}$ s$^{-1}$)}&
\multicolumn{1}{c}{( erg s$^{-1}$)} &
\multicolumn{1}{c}{( erg s$^{-1}$)} \\
\multicolumn{1}{c}{}&
\multicolumn{1}{c}{} &
\multicolumn{1}{c}{} &
\multicolumn{1}{c}{} &
\multicolumn{1}{c}{} &
\multicolumn{1}{c}{} &
\multicolumn{1}{c}{} &
\multicolumn{1}{c}{0.1$-$2.4 keV} &
\multicolumn{1}{c}{0.2$-$4.5 keV}&
\multicolumn{1}{c}{0.1$-$2.4 keV} &
\multicolumn{1}{c}{0.2$-$4.5 keV} \\
\noalign{\smallskip}\hline\noalign{\smallskip}
ROSAT/PSPC     &rp400022n00 &  48536.2 & 0.96 & 0.65 &  16.6 &    936    &  $8.84 \times 10^{-13}$  &   $1.96 \times10^{-12}$   &    $6.43 \times10^{35}$        &    $1.07 \times10^{36}$   \\
ROSAT/PSPC     &rp400022a01 &  48895.7 & 1.87 & 0.66 &   9.0 &    584    &  $1.02 \times 10^{-12}$  &   $2.27 \times 10^{-12}$  &    $7.43 \times10^{35}$        &    $1.24\times10^{36}$    \\
ROSAT/PSPC     &rp400022a02 &  49141.0 & 1.20 & 0.70 &  11.9 &    147    &  $1.95 \times 10^{-13}$  &   $4.32 \times 10^{-13}$ &    $1.42 \times10^{35}$        &     $2.36  \times10^{35}$ \\
XMM/EPIC       &0601212401  &  55011.6 & 0.41 & 0.71 &  25.4 &     58    &      --                 &   $1.55 \times 10^{-14}$ &       --                      &     $7.86 \times10^{33}$\\
\noalign{\smallskip}\hline
\end{tabular}
\tablefoot{
\tablefoottext{a}{Modified Julian date of the beginning of the observation.}
\tablefoottext{b}{Duration of the observation.}
\tablefoottext{c}{Phase = 0.5 corresponds to maximum of optical outbursts.}
\tablefoottext{d}{Unabsorbed luminosity. A source distance of 60 kpc is assumed.}
}
\label{tab:observations}
\end{table*}
%--------------------------

\bexrb\ was detected four times in X-ray observations as listed in Table~\ref{tab:observations}.
The source was found in three ROSAT observations of SMC\,X-1 performed between October 1991 and June 1993 at an off-axis angle of $\sim$27.7\arcmin.
The ROSAT position from \citet[][ their source 460]{2000A&AS..142...41H} is
RA (J2000)=01\rahour23\ramin27\fs2 and Dec (J2000)=$-$73\degr21\arcmin25\arcsec with a 90\% uncertainty of 4.6\arcsec.
This is consistent with the \xmm\ position of the SMC-survey catalogue \citep[][source 3003]{smcsrccat} of
RA (J2000)=01\rahour23\ramin27\fs46 and Dec (J2000)=$-$73\degr21\arcmin23\farcs4 with a $1\sigma$ uncertainty of 1.1\arcsec.

\begin{figure}
  \resizebox{\hsize}{!}{\includegraphics[angle=-90,clip=]{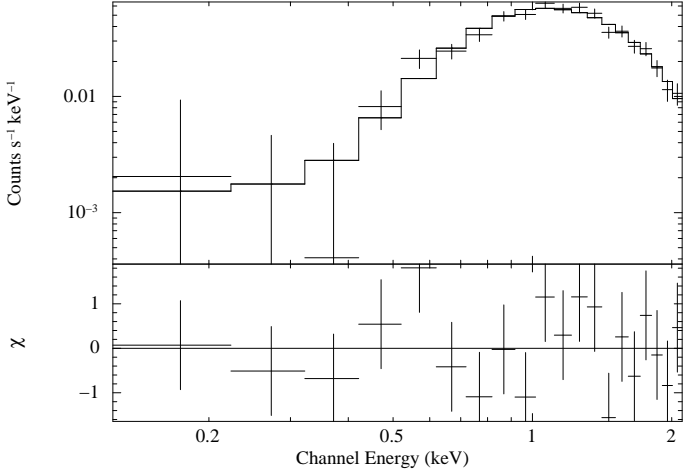}}
  \caption{
    Merged ROSAT/PSPC spectrum of \bexrb\ from observations rp400022n00 and rp400022a01 together with the best-fit power-law model. The lower panel gives the residuals in units of $\sigma$.
  }
  \label{fig:spec}
\end{figure}

ROSAT spectra were extracted using the {\tt ftools} \citep{1995ASPC...77..367B} task {\tt xselect}.
For spectral analysis, we merged the observations rp400022n00 and rp400022a01 to increase the statistics and used {\tt xspec} \citep{1996ASPC..101...17A} version 12.7.0u to fit the spectrum with
a power-law model and photoelectric absorption by a Galactic column density set to N$_{\rm H, gal}=3.3\times10^{20}$ cm$^{-2}$ 
\citep[a foreground \Hone\ map was kindly provided by Erik Muller, see also ][]{2003MNRAS.339..105M}
and an additional column density, accounting for the interstellar medium of the SMC and source intrinsic absorption.
We find a good description of the data by this model with $\chi^2/$dof=14.8/17.
For the SMC and source intrinsic absorption we obtain N$_{\rm H, smc}= 4.32_{-0.28}^{+0.36}\times10^{21}$ cm$^{-2}$.
The total line-of-sight N$_{\rm H, smc}$ column density at the position of \bexrb\ is $3.02\times10^{21}$ cm$^{-2}$ \citep{1999MNRAS.302..417S}, pointing to some intrinsic absorption in the binary system.
The photon index is determined to be $\Gamma=1.18_{-0.54}^{+0.63}$, which suggests a neutron-star BeXRB \citep[$\Gamma \approx 1$, ][]{2004A&A...414..667H}
but we cannot exclude an AGN \citep[$\Gamma \approx 1.75$, ][]{2006A&A...451..457T}.
The ROSAT spectrum together with the best-fit model is presented in Fig.~\ref{fig:spec}.

%--------------------------X-ray spectrum
%--------------------------

The \xmm\ detection is based on only 58 counts.
However, hardness ratios are consistent with a BeXRB nature of the source.
Assuming the same spectral shape for the ROSAT and \xmm\ detections and comparing their fluxes, 
we find a high long-term variability of a factor of 146$\pm$20, that would be atypical for an AGN.

We used a Bayesian odds ratio \citep{1996ApJ...473.1059G,2008A&A...489..327H} and a Rayleigh Z$^2_1$ \citep{1983A&A...128..245B,2002A&A...391..571H} test to search for the spin period of the NS.
The search for pulsations in the ROSAT data with best statistics (rp400022n00) in the (1$-$200) s band did not reveal convincing periodicities.
Note that the ROSAT observation was performed with individual exposures between 200 and 1300 s, preventing the detection of long periods.
The best candidate for the spin period with 1$\sigma$ uncertainty is found at 17.34529(55) s.
Since this period has a significance of 3.3$\sigma$ and could not be confirmed in the other observations, probably owing to the lower number of detected counts, the spin period needs further confirmation.

\section{Analyses and results of optical observations}
\label{sec:optical}

%--------------------------finding chart
\begin{figure}
  \resizebox{\hsize}{!}{\includegraphics[angle=0,clip=]{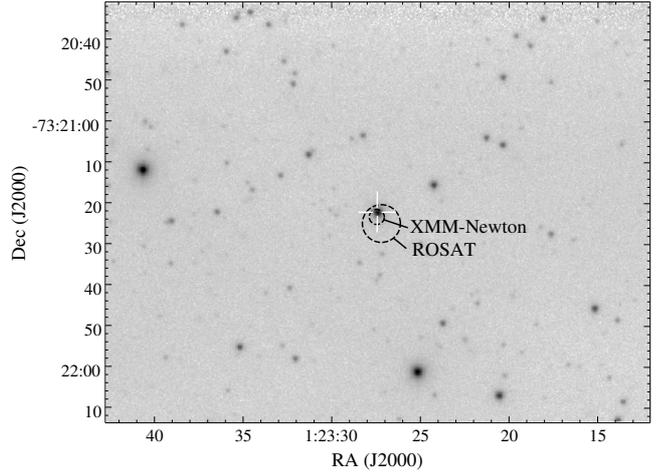}}
  \caption{
    OGLE\,III $I$-band finding chart of \bexrb, marked with a white cross.
    Circles indicate the 90\% position uncertainties of the ROSAT and \xmm\ detections. 
  }
  \label{fig:fc}
\end{figure}
%--------------------------

An $I$-band finding chart from OGLE\,III is presented in Fig.~\ref{fig:fc}.
An optical counterpart, \object{GSC\,09142-02001}, is found with an angular separation to the \xmm\ position of 1.3\arcsec\ for the entry in the photometric catalogue of \citet{2002ApJS..141...81M}, giving 
$U=14.48$ mag,
$B=15.39$ mag,
$V=15.45$ mag, and
$R=15.42$ mag as measured on MJD 51186.
The NIR counterpart, \object{2MASS\,01232743-7321223} \citep{2006AJ....131.1163S}, is found with an angular separation of 1.1\arcsec\ with 
$J=15.32$ mag,
$H=14.82$ mag, and
$K_{\rm S}=15.06$ mag measured on MJD 51034.
The corresponding OGLE sources are OGLE\,III\,SMC121.2.50 and OGLE\,IV\,SMC733.26.24 with a separation of 1.2\arcsec to the X-ray source.  %OGLEIII position: 1:23:27.42 -73:21:22.3
The OGLE $I$-band value of $I\approx15.4$ mag is a typical magnitude of the variable light curve (see below).
The magnitudes and colours are characteristic of a B star in the SMC.
From optical spectroscopy, \citet{2004MNRAS.353..601E} classified the source (\object{2dFS\,3624}) as B1-3\,III star.
Since the density of early-type stars is relatively low in the wing of the SMC, a correlation with the X-ray source by chance is unlikely.

%--------------------------sed
% see sed.pro
\begin{figure}
  \resizebox{\hsize}{!}{\includegraphics[angle=0,clip=]{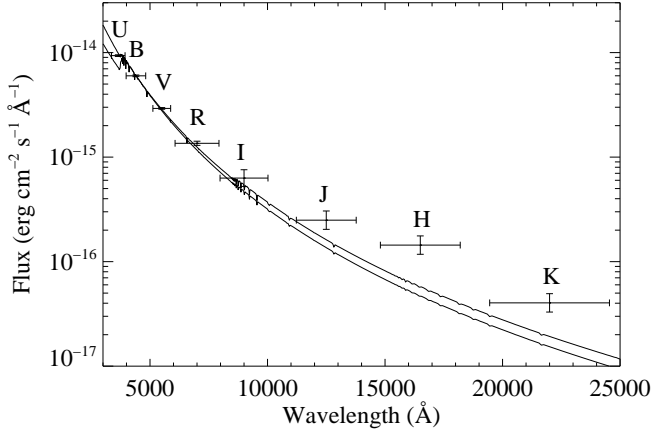}}
  \caption{
Optical and NIR photometry of \bexrb\ dereddened and compared with normalised Kurucz model atmospheres for a B1\,III and B3\,III star. 
Labels give filters.
  }
  \label{fig:sed}
\end{figure}
%--------------------------

We dereddened these magnitudes to correct for a Galactic foreground of $E_{B-V} = 0.07$ mag \citep{1991A&A...246..231S}
and compare them to a Kurucz model atmosphere \citep{1979ApJS...40....1K,2011MNRAS.413.1515H} 
of a B1\,III and B3\,III star ($T_{\rm eff} = 25\,000, 18\,000$ K and $\log{(g)} = 3.0$)
in Fig.~\ref{fig:sed}. 
The models have been normalised to the $B$-band measurement.
To account for possible variability of the source and the non-simultaneous observations, we included an uncertainty of 0.2 mag for the $I$ and NIR magnitudes.
There is evidence for an NIR excess, likely caused by the circum-stellar decretion disc and supporting the Be classification of the star.

%--------------------------light curve
\begin{figure}
  \resizebox{\hsize}{!}{\includegraphics[angle=0,clip=]{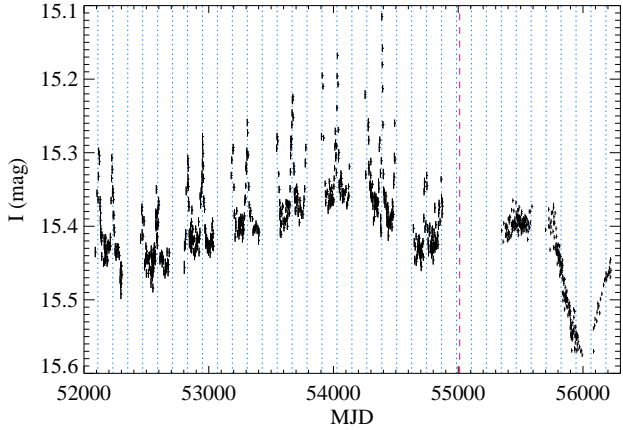}}
  \caption{
    Optical $I$-band light curve of \bexrb\ from OGLE\,III (before MJD 55000) and OGLE\,IV (after MJD 55000).
    Predicted outburst maxima are indicated by the dotted blue lines. 
    The time of the \xmm\ observation is marked by the dashed red line.
  }
  \label{fig:lc}
\end{figure}
%--------------------------LS test
\begin{figure}
  \resizebox{\hsize}{!}{\includegraphics[angle=0,clip=]{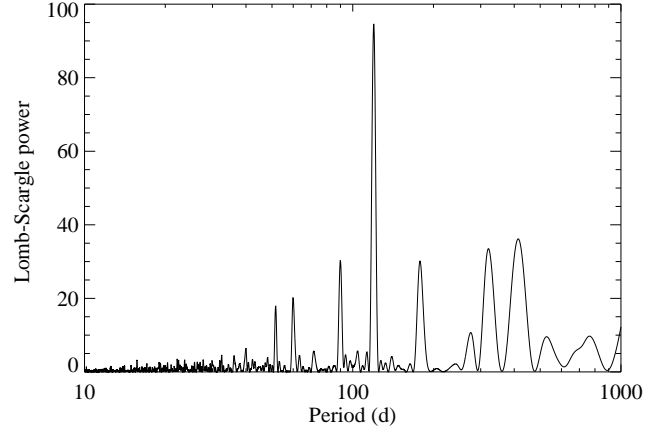}}
  \caption{Lomb-Scargle periodogram of the $I$-band light curve of \bexrb\ from OGLE\,III data before MJD 55000.
      }
  \label{fig:ls}
\end{figure}
%--------------------------light curve convolved
\begin{figure}
  \resizebox{\hsize}{!}{\includegraphics[angle=0,clip=]{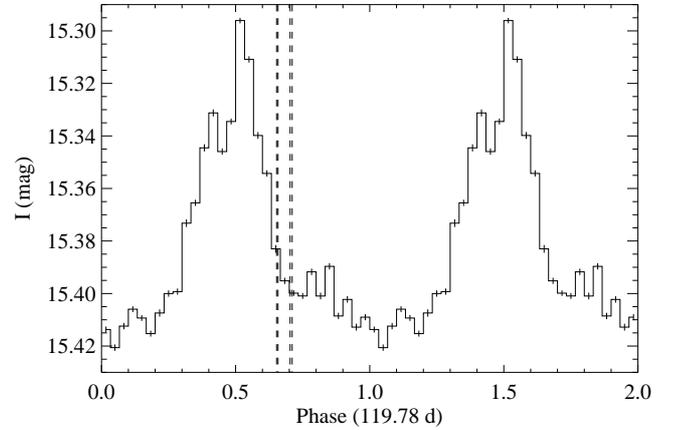}}
  \caption{
    OGLE\,III light curve of \bexrb\ for MJD $<$ 55000 folded with the 119.87 d period.
    Phase = 0.5 corresponds to MJD 54387.14.
    The phases of the dates of the X-ray observations are indicated by dashed lines (see Table~\ref{tab:observations}).
    }
  \label{fig:lc-fold}
\end{figure}
%--------------------------light curve individual outbursts
\begin{figure}
  \resizebox{\hsize}{!}{\includegraphics[angle=0,clip=]{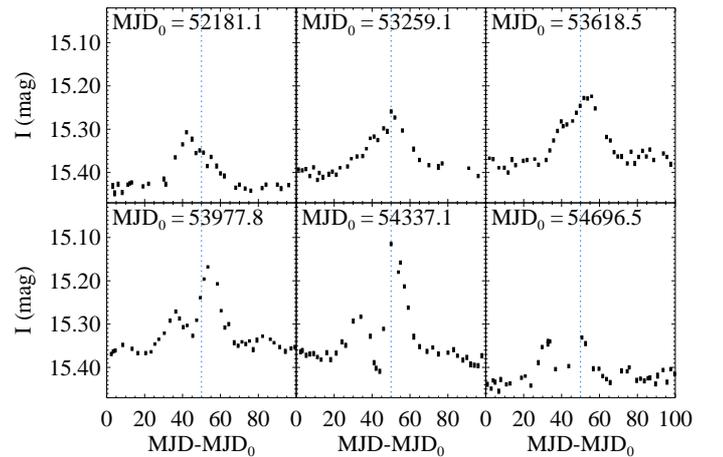}}
  \caption{
    OGLE\,III light curves of \bexrb\ for six individual outbursts.
    Predicted outburst maxima are indicated by the dotted blue lines.
    }
  \label{fig:lc-ind}
\end{figure}

The long-term $I$-band light curve from OGLE phase III and IV is presented in Fig.~\ref{fig:lc}.
Besides a long-term variability, regular outbursts are seen.
The light curve is typical for Be stars.
A Lomb-Scargle periodogram \citep{1976Ap&SS..39..447L,1982ApJ...263..835S} of the OGLE\,III data confirms the variability with a period of 119.87 days (Fig.~\ref{fig:ls}).
We derive an ephemeris for the $I$-band outburst maximum of 
$$ \mathrm{MJD} = 54387.14 \pm 0.10  +  (119.9 \pm 2.5)  \times N  , $$ with $N$ an integer.
The folded $I$-band light curve is presented in Fig.~\ref{fig:lc-fold}.
The average outburst duration (FWHM) is 30 days as derived from the folded light curve. %FWHM is from gaussian-fit sigma
However, the evolution can be different for individual outbursts.
For several outbursts we find a double-peaked shape, as also indicated in the folded light curve.
Other outbursts only show one peak, where the maximum can be $\sim$15 days before or after the expected time of the outburst maximum.
Examples for individual outburst are presented in Fig.~\ref{fig:lc-ind}.
In other BeXRBs, typically a fast rise and exponential decay is observed during optical outbursts \citep{2012MNRAS.423.3663B}.
In general, individual outburst profiles will depend on NS-orbit parameters and perturbations of the decretion disc \citep[][]{1991PASJ...43...75O}.

All X-ray observations were done shortly after the times of possible optical outbursts, as indicated by the vertical lines in Fig.~\ref{fig:lc-fold}.
If both optical and X-ray outburst appear during periastron passage, 
we would expect quasi-simultaneous outbursts as e.g. observed in AX\,J0058-720 \citep{2007A&A...476..317H} and IGR\,J05414-6858 \citep{2012A&A...542A.109S}.
However, we note that the X-ray observations have been performed during times without OGLE coverage where the occurrence of optical outbursts is uncertain.
The recently observed decrease in the $I$-band light curve and the vanishing of the outbursts can be caused by a loss of the decretion disc around the Be star.
Interestingly, first the outbursts were missing after MJD 55400, although the quiescent  $I$-band flux was at a similar level when outbursts have been observed around MJD 53000.
Between MJD 55700 and 56000, a drop of the $I$-band flux was observed, followed by a brightening, probably due to a recovering of the decretion disc.

\begin{figure}
  \resizebox{\hsize}{!}{\includegraphics[angle=0,clip=]{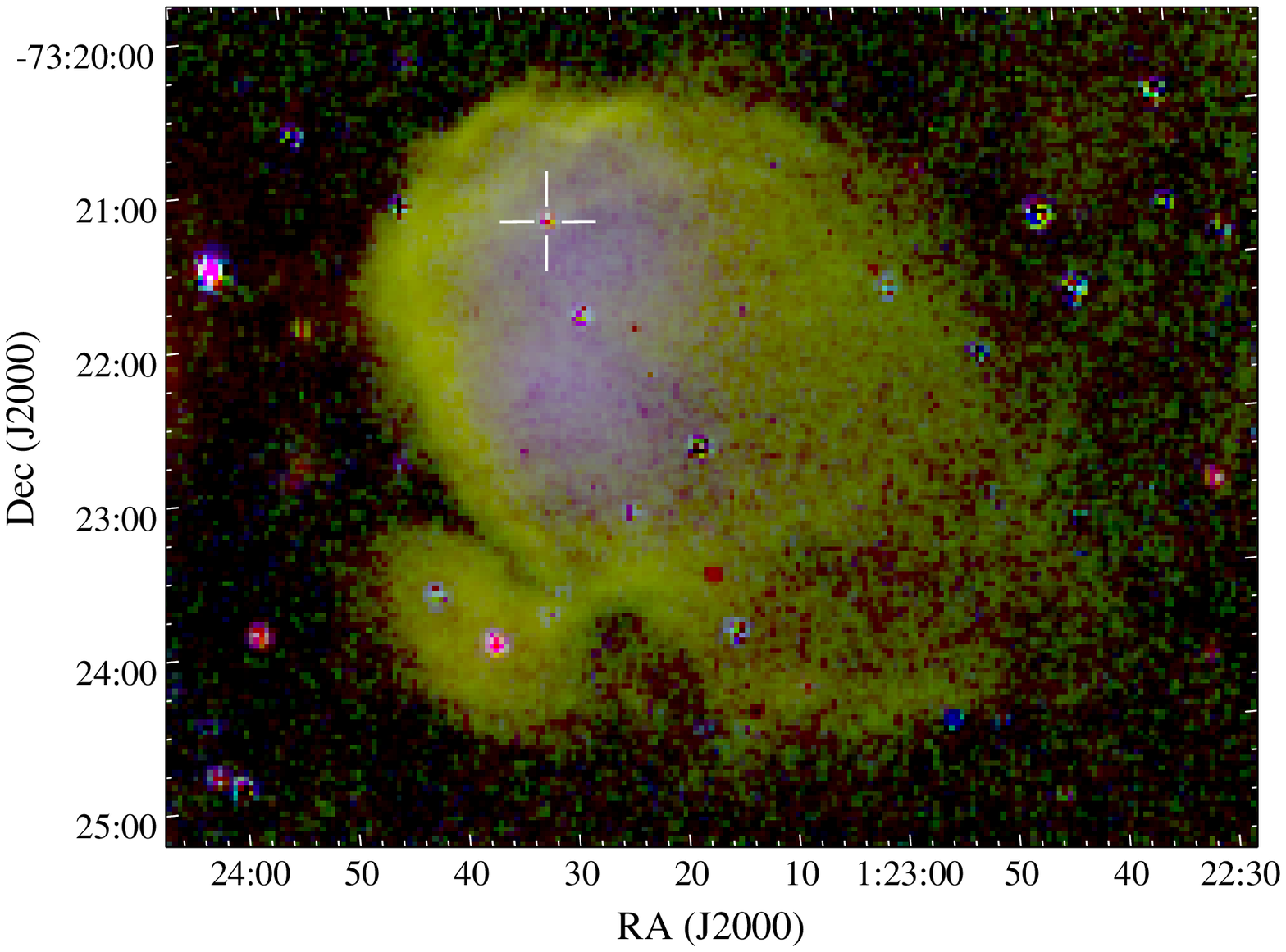}}
  \caption{Continuum-subtracted emission-line image of DEM\,S\,160 from MCELS \citep[][]{2005AAS...20713203W}.
  Red/green/blue give logarithmically scaled intensities for \Halp/[\ion{S}{ii}]/[\ion{O}{iii}].
  \bexrb\ is marked with a white cross.
      }
  \label{fig:mcels}
\end{figure}

We note that \bexrb\ is along the line of sight to the 
\ion{H}{ii} region DEM\,S\,160 \citep{1976MmRAS..81...89D}, see Fig.~\ref{fig:mcels}.
However, due to the high angular separation of $\sim$1\arcmin\ to the centre of DEM\,S\,160, 
which corresponds to a distance of at least 17.5 pc, both sources are likely not connected directly to each other
but might have formed during the same star-formation event.

%--------------------------MCELS
%--------------------------

\section{Discussion and conclusions}
\label{sec:conclusion}

We analysed ROSAT and OGLE data of \bexrb.
Due to the high X-ray variability by a factor of at least $\sim$150 and the properties of the optical counterpart, we identify \bexrb\ as new BeXRB in the wing of the SMC.
The $I$-band light curve of the optical counterpart shows a typical behaviour for a Be star in a BeXRB.
The periodic outbursts are likely caused by binarity and reveal the orbit period of the NS of $P_{\rm orb} =119.78$ d.
This places \bexrb\ in the sample of long-orbit systems \citep{2011Natur.479..372K}.
From the Corbet relation \citep{1984A&A...141...91C,2009IAUS..256..361C}, we can roughly estimate the spin period of the neutron star to be between 10 and 800 s.
Therefore, the source is unlikely to be identified with the 2.16 s pulsar XTE\,J0119-731 \citep{2003IAUC.8064....4C} that was found by RXTE with an angular separation of 18\arcmin.
No indication of this period was found in the X-ray periodograms of \bexrb.
Further X-ray observations are necessary of the system in outburst, to confirm the possible spin period of 17.3 s and the neutron-star nature of the compact object.
If the current low-state of the $I$-band emission is caused by a loss of the decretion disc, no X-ray activity is expected.
Follow-up X-ray observations are most likely successful, during new $I$-band outbursts, which indicate interaction of the NS and the decretion disc.
Optical outbursts might reappear with brightening $I$-band emission and will be found in future OGLE monitoring.
The system is included in the X-Ray variables OGLE Monitoring \citep[XROM,][]{2008AcA....58..187U} system.

\begin{acknowledgements}
The OGLE project has received funding from the European Research Council
under the European Community's Seventh Framework Programme
(FP7/2007-2013) / ERC grant agreement no. 246678 to AU.
RS acknowledges support from the BMWI/DLR grant FKZ 50 OR 0907.
\end{acknowledgements}

\bibliographystyle{aa}
\bibliography{../auto,../general}

\end{document}